Research article

Open Access

# Electronic structure, linear, nonlinear optical susceptibilities and birefringence of $CuInX_2$ (X = S, Se, Te) chalcopyrite-structure compounds

Ali Hussain Reshak*[1] and S Auluck[2]

Address: [1]Institute of Physical Biology, South Bohemia University, Institute of System Biology and Ecology-Academy of Sciences, Nove Hrady 37333, Czech Republic and [2]Physics Department, Indian Institute of Technology, Kanpur (UP) 208016, India

Email: Ali Hussain Reshak* - maalidph@yahoo.co.uk; S Auluck - sauluck@iitk.ac.in

* Corresponding author

## Abstract

The electronic structure, linear and nonlinear optical properties have been calculated for $CuInX_2$ (X = S, Se, Te) chalcopyrite-structure single crystals using the state-of-the-art full potential linear augmented plane wave (FP-LAPW) method. We present results for band structure, density of states, and imaginary part of the frequency-dependent linear and nonlinear optical susceptibilities. We find that these crystals are semiconductors with direct band gaps. We have calculated the birefringence of these crystals. The birefringence is negative for $CuInS_2$ and $CuInSe_2$ while it is positive for $CuInTe_2$ in agreement with the experimental data. Calculations are reported for the frequency-dependent complex second-order non-linear optical susceptibilities $\chi^{(2)}_{123}(\omega)$. The intra-band and inter-band contributions to the second harmonic generation increase when we replace S by Se and decrease when we replace Se by Te. We find that smaller energy band gap compounds have larger values of $\chi^{(2)}_{123}(0)$ in agreement with the experimental data and previous theoretical calculations.

## 1. Introduction

The ternary $A^{XI}B^{XIII}C_2^{XVI}$ semiconducting compounds which crystallize in chalcopyrite structure have drawn much attention in recent years [1-3,31,66]. They form a large group of semiconducting materials with diverse optical, electrical, and structural properties [4-11]. Ternary chalcopyrite appear to be promising candidates for solar-cells applications [12,64,65], light-emitting diodes [13], nonlinear optics [14], and optical frequency conversion applications in all solid state based tunable laser systems. They have potentially significant advantages over dye lasers because of their easier operation and the potential for more compact devices. Tunable frequency conversion in mid infrared (IR) is based on optical parametric oscillators (OPO) using pump lasers in near IR [15]. On the other hand frequency doubling devices also allow one to expand the range of powerful lasers to far IR such as $CO_2$ lasers [15-17].

In last decade, first-principle calculations have been successfully used to obtain different properties of materials. The structural parameters and dynamical properties of crystals determine a wide range of microscopic and macroscopic behavior: diffraction, sound velocity, elastic constants, Raman and infrared absorption, inelastic neutron scattering, specific heat, etc. The $A^{XI}B^{XIII}C_2^{XVI}$ chalcopyrites concerned in this paper use copper as the group *XI* element, indium as the group *XIII* element, and sulphur, selenium and tellurium as the group *XVI* element. Since these compounds display large birefringence [18], they are potentially interesting as nonlinear optical materials, as well as semiconductors. $CuInTe_2$ has positive birefringence whereas $CuInS_2$ and $CuInSe_2$ have negative birefringent [19,20]. So far, however, the trends of the coupling coefficients in these materials are not well understood. Rashkeev *et al.* [17] used the linear muffin-tin orbitals (LMTO) method within muffin-tin approximation (MTA) to predict enhancement of $\chi^{(2)}$ $(\omega)$ by substitution of S by Se and Te. It is well known that MTA works reasonably well in highly coordinated systems, such as face-centered cubic (FCC) metals [56]. For covalently bonded solids or layered structures, such as chalcopyrites, the MTA is a poor approximation and leads to discrepancies with experiments [52]. The more general treatment of the potential, such as provided by a full potential (FP) method has none of the drawbacks of the atomic sphere approximation (ASA) and MTA based methods. In full potential methods the potential and charge density are expanded into lattice harmonics inside each atomic sphere and as a Fourier series in the interstitial region. In the present work we use the full potential linear augmented plane wave (FP-LAPW) method which has proven to be one of the most accurate methods [54,55] for computation of the electronic structure of solids within density functional theory (DFT). Hence the effect of full potential on the linear and nonlinear optical properties can be ascertained.

The dynamical properties of chalcopyrites are well known and have been described by many workers. There is much research interests in their hardness [21-23] and pressure-induced behaviors [24,25]. Recently Eryigit et al. [53] have performed a first principles study of structural, dynamical, and dielectric properties of the chalcopyrite semiconductor $CuInS_2$. Yet up to now there is no comprehensive work that concerns the electronic structure, frequency dependent dielectric function, birefringence and frequency dependent nonlinear optical (NLO) properties of these compounds, although their potential NLO applications has been emphasized. A compre-hensive understanding of the origin of the optical nonlinearity and high $\chi^{(2)}(\omega)$ of these materials are very interesting subjects.

There exist a number of calculations of electronic band structure and optical properties using different methods starting from Miller's empirical rule to the current first principles [26-31] methods. We are not aware of any full-potential calculations for these compounds. Most the existing *ab initio* calculations are based on MTA. We therefore think it worthwhile to perform *ab initio* calculations using a full potential method. The calculated energy gaps vary from 0.01 to 0.812 eV for $CuInS_2$, 0.01 to 0.416 eV for $CuInSe_2$ and 0.18 to 0.424 eV for $CuInTe_2$. Thus there is a large variation in the energy gaps, suggesting that the energy band gap depends on the method of band structure calculation. Also some of the calculated energy gaps are equal to the measured energy gap which is not expected from calculations based on the local density approximation. There is a dearth of theoretical calculations for the frequency dependent dielectric function $\varepsilon(\omega)$. Also there is no experimental data for the NLO. There is only one report [31] in which they have analyzed the second order of NLO from the chemical bond viewpoint. We think it is worthwhile to perform these calculations. Our calculations will highlight the effect of replacing S by Se and Se by Te on the electronic and optical properties in the $CuInX_2$ compounds. Our motivation in this paper is to understand the origin of the high $\chi^{(2)}(\omega)$ and the degree of birefringence in these materials as well as to study the trends with moving from S to Se to Te.

## 2. Details of Calculations

The ternary $CuInX_2$ (X = S, Se, Te) semiconducting compounds crystallize in the chalcopyrite structure with tetragonal space group $I\bar{4}2d\left(D_{2d}^{12}\right)$ having four formula units in each unit cell. $CuInX_2$ is a ternary analog of diamond structure and essentially a superlattice (or superstructure) of zinc blende. Like the atoms in diamond and zinc blende structures, each constituent atom in these ternary compounds, *XI*, *XIII*, and *XVI*, is tetrahedrally coordinated to four neighboring atoms. The Cu atom is located at (0, 0, 0), (0, 0.5, 0.25), In atom at (0, 0, 0.5), (0, 0.5, 0.75), and X atoms at ($x$, 0.4, 0.125), (-$x$, 0.75, 0.125), (0.75, $x$, 0.875), (0.4,- $x$, 0.875), where $x$ is equal to 0.20, 0.22, 0.225 for $CuInS_2$, $CuInSe_2$, $CuInTe_2$ respectively. In the present work we used the experimental lattice parameters [32] as listed in Table 1. These ternary compounds have a small energy gap and this proved to be of great interest in the nonlinear optical properties [33,18].

In our calculations we use the state-of-the-art full potential linear augmented plane wave (FPLAPW) method in a scalar relativistic version as embodied in the WIEN2k code [34]. This is an implementation of the DFT with different possible approximations for the exchange-correlation (XC) potential. Exchange and correlation are treated within the local-density approximation (LDA), and scalar relativistic equations are used to obtain self-consistency. Kohn-Sham equations are solved using a basis of linear APW's. The LDA is known to understand energy gaps by up to 40%. Attempts to improve on this by using LDA+U, SIC (self interaction methods) and *GW* methods. These methods are very time consuming and are a challenge by itself. As the calculation of the nonlinear optical properties are very time consuming. Hence we have taken a simple approach of using LDA and simply correcting the energy by a rigid shift of the valence/conduction bands.

**Table 1: Lattice parameters, energy gaps, our calculated $\varepsilon_1^{\perp}(0)$ and $\varepsilon_2^{\|}(0)$ and Δ*n*(0).**

| | $CuInS_2$ | $CuInSe_2$ | $CuInTe_2$ |
|---|---|---|---|
| A (Å) | 5.52[a] | 5.78[a] | 6.179[a] |
| C (Å) | 11.08[a] | 11.57[a] | 12.36[a] |
| $E_g^{exp.}$ (eV) | 1.53[b, r], 1.55[e], 1.48[i], 1.54[d] | 1.04[b], 0.96[g], 1.01[h], 0.98[i, r], 0.88[d] | 1.06[b, j], 0.96[f] |
| $E_g^{Theory}$ (eV) | 0.01[c], -0.14[r], 0.812* | 0.01[c], -0.2[r], 0.416* | 0.18[c], 0.424* |
| $\varepsilon_1^{\perp}(0)$ | 11.3*, 9.71[P], 8.6[q] | 14.8* | 13.5* |
| $\varepsilon_1^{\|}(0)$ | 11*, 9.68[P], 8.4[q] | 14.6* | 13.8* |
| Δ*n*(0) | -0.018[k], -0.019* | -0.012[m], -0.013* | 0.002* |
| $N_0$ | 2.532[n, o], 2.71* | 2.606[n], 2.83* | 2.720[n], 2.91* |

[a]Ref.32 [b]Ref.36 [c]Ref.17 [d]Ref.37 [e]Ref.38 [f]Ref. 11 [g]Ref.39 [h]Ref.40 [i]Ref.41 [j]Ref.42 [k]Ref.19 [m]Ref.20 [n]Ref.31 [o]Ref.57 [P]Ref.53 [q]Ref.62 [r]Ref.63
*This work

In order to achieve energy eigenvalues convergence, the wave functions in the interstitial region were expanded in plane waves with a cut-off $K$max = $9/R_{MT}$, where $R_{MT}$ denotes the smallest atomic sphere radius and $K$max gives the magnitude of the largest $K$ vector in the plane wave expansion. The $R_{MT}$are taken to be 2.00 atomic units (a.u.) for Cu, In, S, Se and Te respectively. The valence wave functions in side the spheres are expanded up to $l_{max}$ = 10 while the charge density was Fourier expanded up to $G_{max}$ = 14. Self-consistent calculations are considered to be converged when the total energy of the system is stable within $10^{-4}$ Ry. The integrals over the Brillouin zone are performed using 250 ***k***-points in the irreducible Brillouin zone (IBZ). The BZ integrations are carried out using the tetrahedron method [67,68]. The frequency dependent linear optical properties are calculated using 500 ***k***-points and the nonlinear optical properties using 1500 $\vec{k}$ points in the IBZ. Both the plane wave cutoff and the number of ***k***-points were varied to ensure total energy convergence.

## 3. Results and Discussion

### *3.1. Band structure and density of states*

The band structure, total density of states (TDOS) along with the X-p/s, and Cu-d/p/s partial DOS for $CuInX_2$ (X = S, Se, Te) compounds are shown in Figure 1. In all cases, the valence band max-imum (VBM) and the conduction band minimum (CBM) are located at Γ resulting in a direct energy gap of 0.812, 0.416, and 0.424 eV for $CuInS_2$, $CuInSe_2$ and $CuInTe_2$, respectively. $CuInSe_2$ and $CuInTe_2$ have a reduction of the gap in comparison to $CuInS_2$. The reduction in the energy gap can be attributed due to the fact that the conduction bands shift towards Fermi energy ($E_F$) when we move from S to Se to Te. In the conduction bands shifting Cu-s states has a small effect whereas shifting In-s states has a strong effect in opening the energy band gap, while leaving the valence bands unchanged. The overall reduction in the energy band gap is consistent with an overall weakening of the bonds, and, therefore, with a smaller bonding antibonding splitting. A comparison of the experimental and theoretical band gaps is given in Table 1. The calculated energy gaps are smaller than the experimental gaps as expected from an LDA calculation [17]. We note that the energy gap decreases when S is replaced by Se/Te in agreement with the experimental data. This trend in the energy band gaps is not present in previous LMTO calculations [17], suggesting that the MTA and ASA are poor approximations and lead to discrepancies on comparison with the experimental data [52]. The band structure and hence the DOS can be divided into six groups. From the PDOS we are able to identify the angular momentum character of the various structures. The lowest group has mainly In-d states. The second group between -11.0 to -14.0 eV has significant contributions from X-s states. The third group -6.0 to -7.0 eV is mainly In-X bond. The groups from -5.5 eV up to Fermi energy ($E_F$) are due to Cu-d states with some contribution from X-p states. The electronic structure of the upper valence band is dominated by Cu-d and X-p states. We note that most of Cu-d character is concentrated in the upper valence band. The Cu-s states are pushed from the valence bands into conduction bands. The last group from 0.5 eV and above has contributions from X-p, Cu-spd, and In-sp states. The trends in the band structures (as we move from S to Se to Te) can be summarized as follows: (1). The second group in $CuInSe_2$ is shifted towards lower energies by around 0.5 eV in comparison with $CuInS_2$, while in $CuInTe_2$ it is shifted towards higher energies by around 1 eV, with reduces the bandwidth of both $CuInSe_2$ and $CuInTe_2$ with respect to $CuInS_2$. (2). The bandwidth of third group is increased with moving from S to Se to Te, this group is shifted towards higher energies by around 0.2 eV. (3). The bandwidth of conduction band reduces slightly by around 0.5 eV on going from S to Se to Te causing to increase the gap between the third and fourth groups. From the PDOS, we note a strong hybridization between Cu-d and X-p states around -4 eV. Following Yamasaki et al. [43] we can define degree of hybridization by the ratio of Cu-d states and X-p states within the muffin tin sphere. Based on this we can say that the hybridization between Cu-d and X-p states becomes weak when we move from S to Se to Te. Also we note that Cu-s strongly hybridized with Cu-p, and In-s with In-p states. Cu-s, Cu-p, In-s, and In-p show strong hybridization with In-d states.

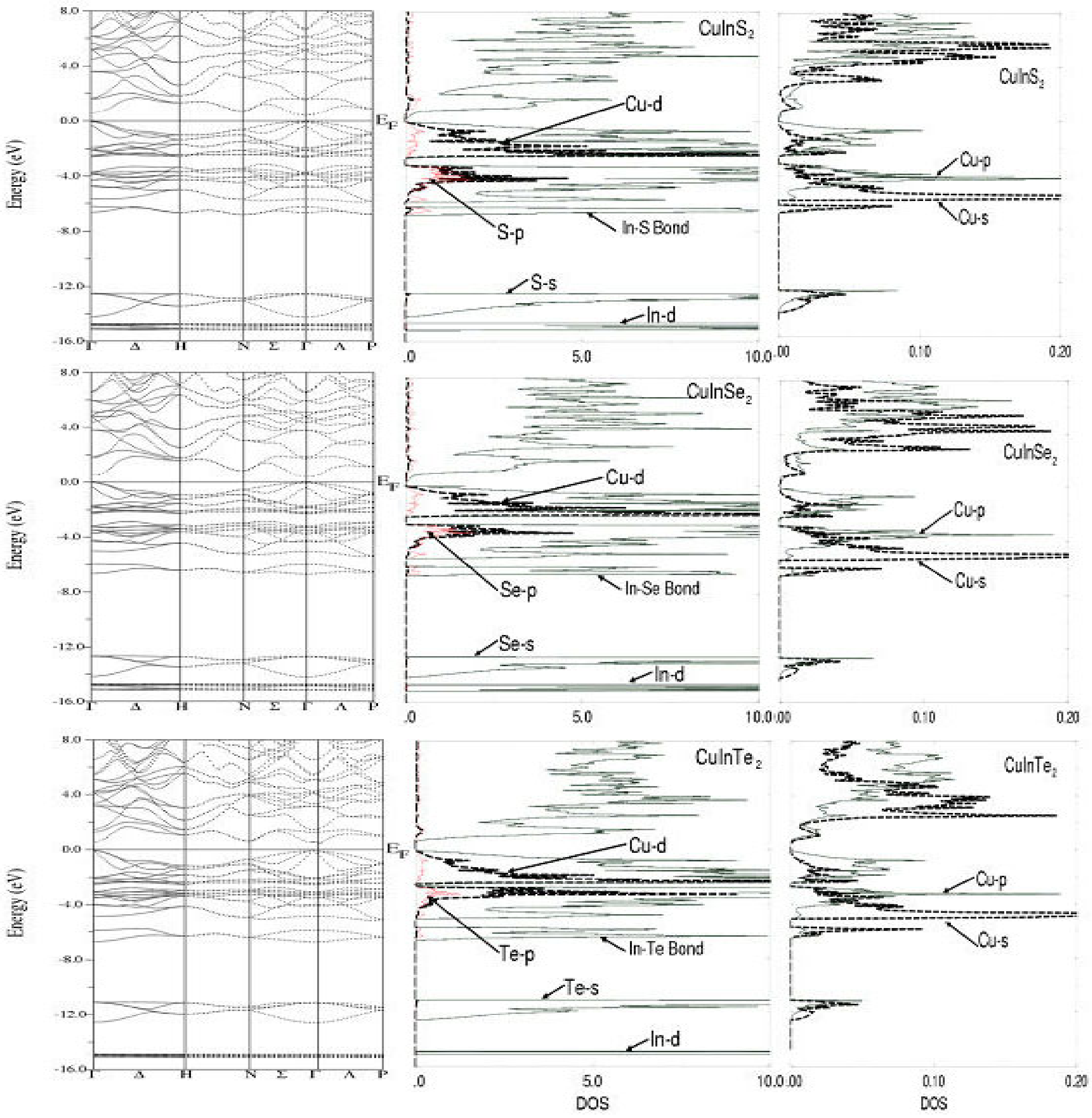

**Figure 1**
Band structure and total density of states (states/eV-unit cell), along with Cu-p/s, X-p, and Cu-d partial densities of states.

### 3.2. Linear optical response and birefringence

Since the investigated compounds have tetragonal symmetry, we need to calculate two dielectric tensor components, corresponding to electric field $\vec{E}$ perpendicular and parallel to c-axis, to completely characterize the linear optical properties. These are $\varepsilon_2^{\perp}(\omega)$ and $\varepsilon_2^{\parallel}(\omega)$, the imaginary parts of frequency dependent dielectric function. We have performed calculations of the frequency dependent dielectric function for these compounds using the expressions [44,45]

$$\varepsilon_2^{\parallel}(\omega) = \frac{12}{m\omega^2} \int_{BZ} \sum \frac{\left| P_{nn'}^{Z}(k) \right|^2 dS_k}{\nabla \omega_{nn'}(k)}$$

$$\varepsilon_2^{\perp}(\omega) = \frac{6}{m\omega^2} \int_{BZ} \sum \frac{\left[ \left| P_{nn'}^{X}(k) \right|^2 + \left| P_{nn'}^{Y}(k) \right|^2 \right] dS_k}{\nabla \omega_{nn'}(k)}$$

The above expressions are written in atomic units with $e^2 = 1/m = 2$ and $\hbar = 1$. where $\omega$ is the photon energy and $P_{nn}^{X}(k)$ is the x component of the dipolar matrix elements between initial $|nk\rangle$ and final $|n'k\rangle$ states with their eigenvalues $E_n(k)$ and $E_{n'}(k)$, respectively. $\omega_{nn'}(k)$ is the energy difference $\omega_{nn'}(k) = E_n(k) - E_{n'}(k)$ and $S_k$ is a constant energy surface $S_k = \{k; \omega_{nn'}(k) = \omega\}$

Figure 2 shows the calculated imaginary part of the anisotropic frequency dependent dielectric function $\varepsilon_2^{\perp}(\omega)$ and $\varepsilon_2^{\parallel}(\omega)$. A considerable anisotropy is found between $\varepsilon_2^{\perp}(\omega)$ and $\varepsilon_2^{\parallel}(\omega)$. Broadening is taken to be 0.02 eV. It is well known that LDA calculations underestimate the energy gaps. A very simple way to overcome this drawback is to use the scissors correction, which merely makes the calculated energy gap equal to the experimental gap. Our calculated optical properties are scissors corrected [14,35] by 0.716, 0.622, and 0.635 eV for $CuInS_2$, $CuInSe_2$, and $CuInTe_2$ respectively. These values are the difference between the calculated and measured energy gap (Table 1). We note that the magnitude of $\varepsilon_2(\omega)$ is increased when move from S to Se to Te. Our calculated $\varepsilon_2^{\perp}(\omega)$ and $\varepsilon_2^{\parallel}(\omega)$ show reasonable agreement with the experimental data [20].

It would be worthwhile to attempt to identify the transitions that are responsible for the structures in $\varepsilon_2^{\perp}(\omega)$ and $\varepsilon_2^{\parallel}(\omega)$ using our calculated band structure. Generally the peaks in the optical response are caused by the electric-dipole transitions between the valence and conduction bands. These peaks in the linear optical spectra can be identified from the band structure. Figure 3, show the transitions which are responsible for the structures in $\varepsilon_2^{\perp}(\omega)$ and $\varepsilon_2^{\parallel}(\omega)$. In order to identify these structures we need to look at the optical matrix elements. We mark the transitions, giving the major structure in $\varepsilon_2^{\perp}(\omega)$ and $\varepsilon_2^{\parallel}(\omega)$ in the band structure diagram. These transitions are labeled according to the optical transitions in Figure 2. For simplicity we labeled the transitions in Figure 3, as A, B, and C. The transitions (A) are responsible for the structures of $\varepsilon_2^{\perp}(\omega)$ and $\varepsilon_2^{\parallel}(\omega)$ in the energy range between 0.0 eV and 2 eV, the transitions (B) are respon-

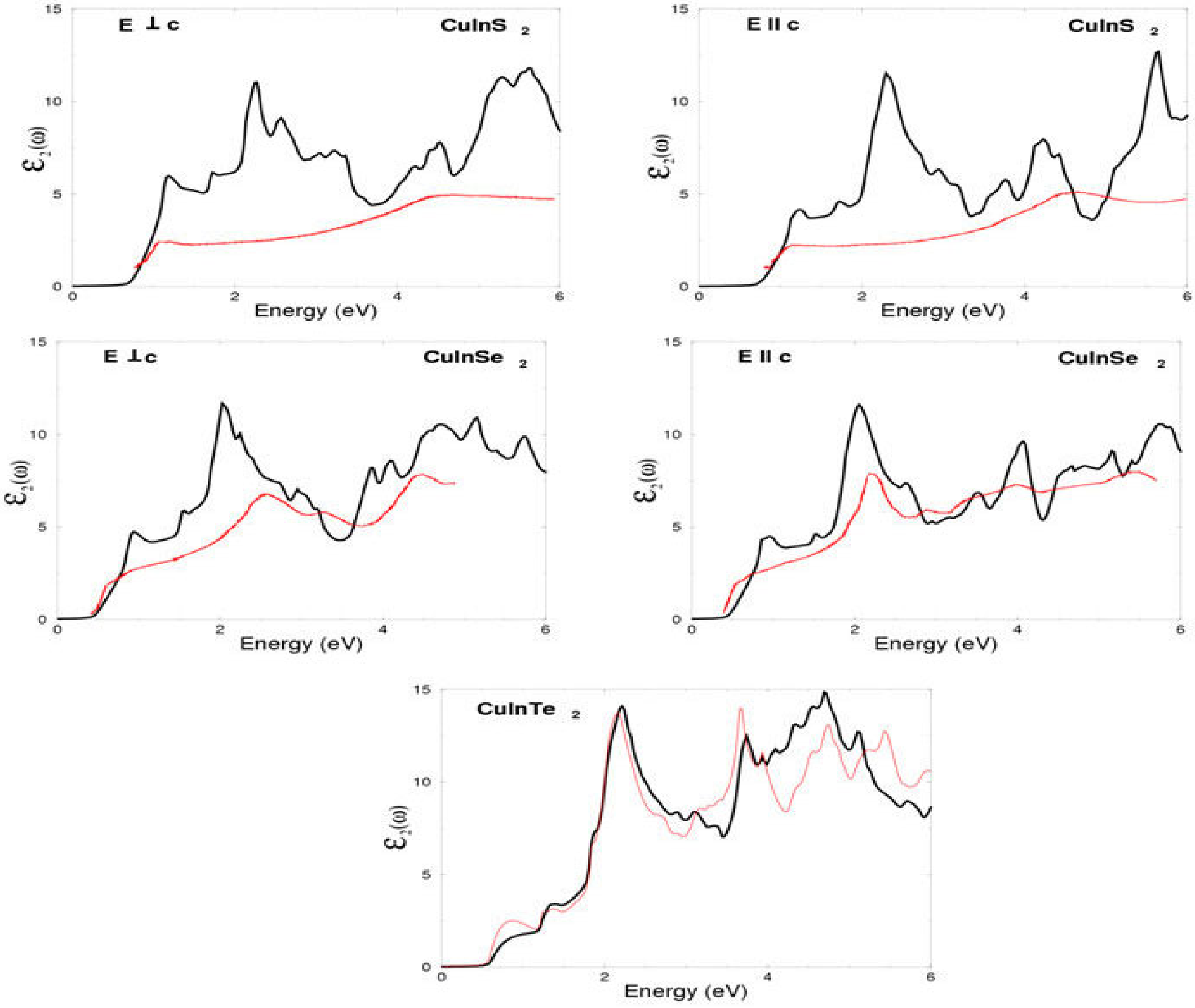

**Figure 2**

Calculated $\varepsilon_2^{\perp}(\omega)$ and $\varepsilon_2^{\parallel}(\omega)$ (**dark curve**) along with the experimental data[20] (light curve) for $CuInS_2$ and $CuInSe_2$, and $\varepsilon_2^{\perp}(\omega)$ (**dark curve**) and $\varepsilon_2^{\parallel}(\omega)$ (light curve) for $CuInTe_2$.

sible for the structures of $\varepsilon_2^{\perp}(\omega)$ and $\varepsilon_2^{\parallel}(\omega)$ in the energy range 2.0 eV and 4.0 eV, and the transitions (C) are responsible for the structures of $\varepsilon_2^{\perp}(\omega)$ and $\varepsilon_2^{\parallel}(\omega)$ between 4.0 eV and 6.0 eV.

The calculated $\varepsilon_1^{\perp}(0)$ and $\varepsilon_2^{\parallel}(0)$ is listed in Table I. We note that a smaller energy gap yields a larger $\varepsilon_1^{\perp}(0)$ and $\varepsilon_2^{\parallel}(0)$ value. This could be explained on the basis of the Penn model [46], where $\varepsilon(0)$ is related to $E_g$ by the equation $\varepsilon(0) \approx 1 + (\hbar\omega_P/E_g)^2$. Hence smaller $E_g$ yields a larger $\varepsilon(0)$.

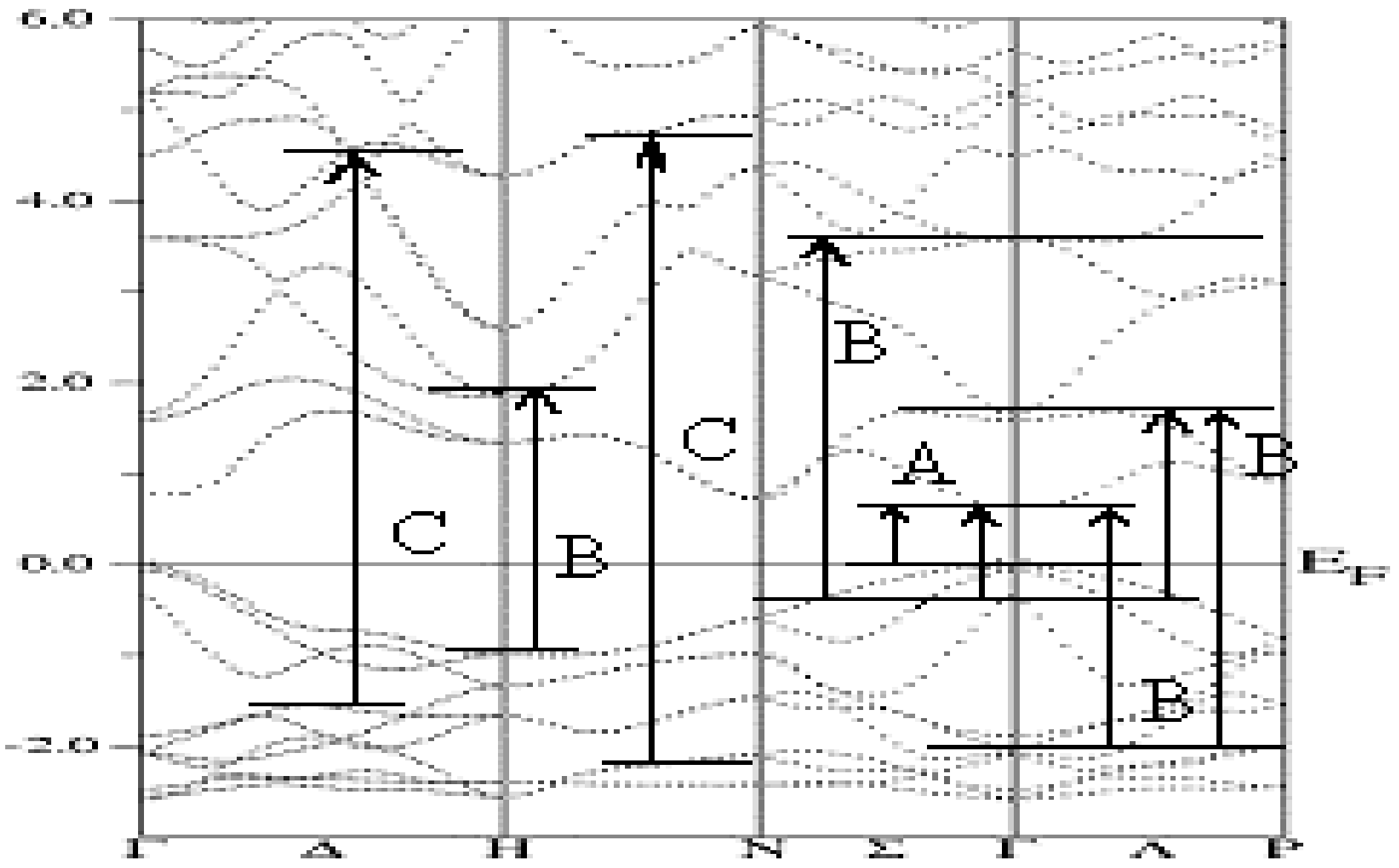

**Figure 3**
The optical transitions depicted on a generic band structure.

These compounds show considerable anisotropy which is lead to an important quantity in SHG and OPO. The important quantity is the birefringence. The birefringence is important to fulfill phase-matching condition. The birefringence can be calculated from the linear response functions from which the anisotropy of the index of refraction is obtained. Using the expression [47]

$$n(\omega) = \left(1/\sqrt{2}\right)\left[\sqrt{\varepsilon_1(\omega)^2 + \varepsilon_2(\omega)^2} + \varepsilon_1(\omega)\right]^{1/2}$$

we can determine the value of the extraordinary and ordinary refraction indices. The birefringence is the difference between the extraordinary and ordinary refraction indices, $\Delta n = n_e - n_0$, where $n_e$ is the index of refraction for an electric field oriented along **c**-axis and $n_0$ is the index of refraction for an electric field perpendicular to **c**-axis.

Figure 4 shows the birefringence $\Delta n(\omega)$ for $CuInX_2$ compounds. The birefringence is important only in the non-absorbing region, which is below the gap. In general we note that the shape of $\Delta n(\omega)$ for the three compounds is rather similar. This attributed due to the fact that these compounds have similar band structures. This curve shows strong oscillations around zero in the energy range up to 12.5 eV. Thereafter it drops to zero. We find that the birefringence is negative for $CuInS_2$ and $CuInSe_2$ and positive for $CuInTe_2$ in agreement with the experimental data [19,20] (Table 1). We have compared our calculated refractive index with that obtained by D. Xue et al. [31] in Table 1. Good agreement is found. This agreement will be in the favor of our calculated $\chi^{(2)}_{ijk}(\omega)$. Previous work of D. Xue et al. [58], discussed the correlation between the nonlinear tensor *dij* and refractive index ($n_0$). They compared the theoretical predictions of the nonlinear optical tensor coefficient $d_{11}$ of $K_4Gd_2(CO_3)_3Fe_4$ with the experimental data and found reasonable agreement with the experimental data of Mercier et al. [59]. This means that there is a correlation between the values of the refractive indices ($n_0$) and nonlinear tensor coefficient *dij*. Based on these results we can say that if our calculated refractive indices ($n_0$) are in good agreement with that obtained by D. Xue et al. [31], then we expect our nonlinear optical susceptibilities $\chi^{(2)}_{312}(0)$ to be in good agreement with the nonlinear tensor coefficient $d_{36}$ which is obtained by D. Xue et al. [31].

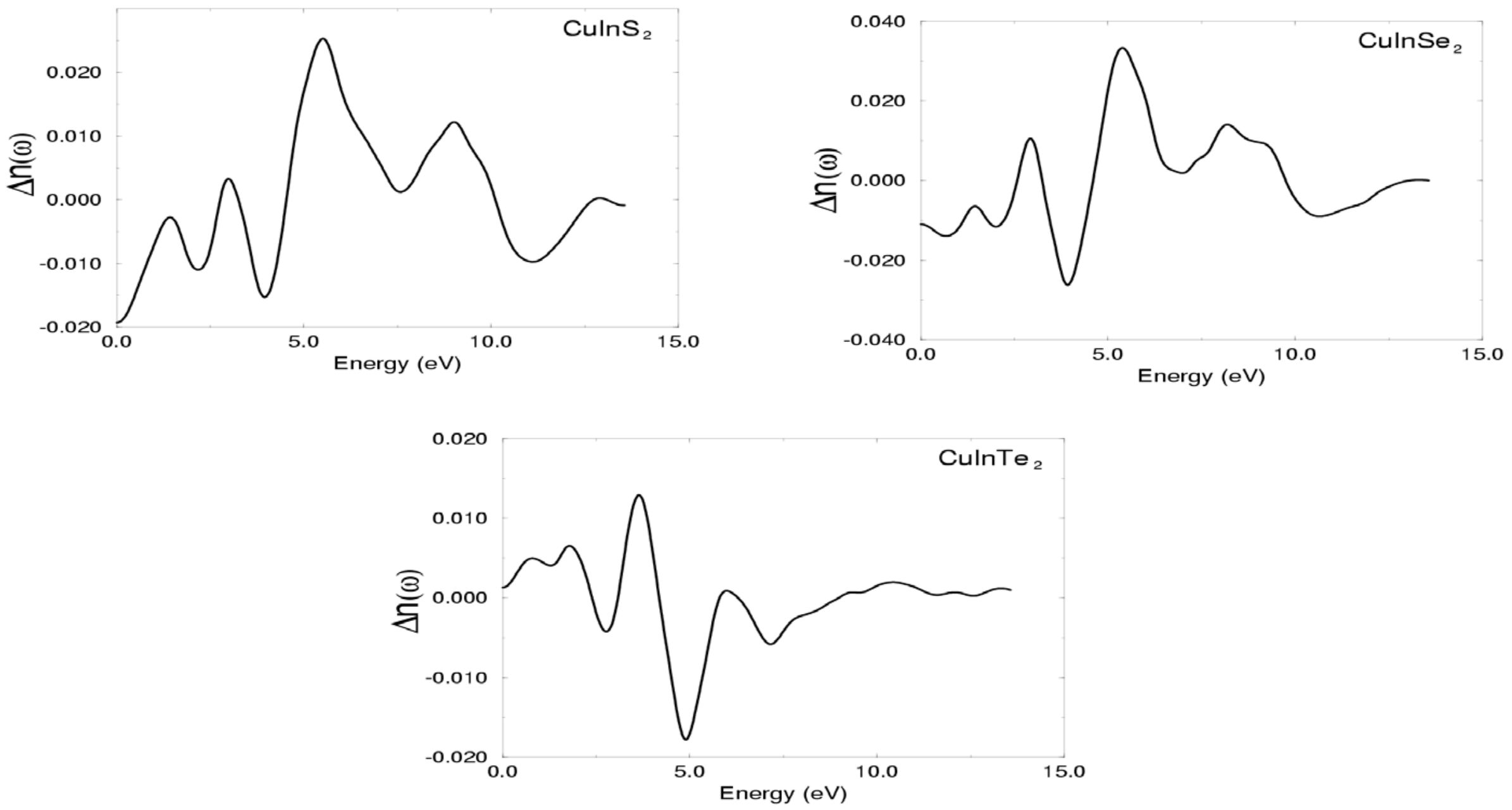

**Figure 4**
Calculated $\Delta n(\omega)$ for $CuInX_2$ compounds.

### 3.3. Non-linear optical response

The complex second-order nonlinear optical susceptibility tensor $\chi^{(2)}_{abc}(-2\omega;\omega;\omega)$ can be generally written as the sum of three physically different contributions in the form [48]:

$$\chi^{abc}_{\text{inter}}(-2\omega;\omega;\omega) = \frac{e^3}{\hbar^2\Omega}\sum_{cvn,k}\frac{r^a_{vc}\left\{r^b_{cn}r^c_{nv}\right\}}{(\omega_{nv}-\omega_{cn})}\left[\frac{2f_{vc}}{\omega_{cv}-2\omega}+\frac{f_{nc}}{\omega_{nc}-\omega}+\frac{f_{vn}}{\omega_{vn}-\omega}\right]$$

$$\chi^{abc}_{\text{intra}}(-2\omega;\omega;\omega) = \frac{ie^3}{2\hbar^2\Omega}\sum_{cv,k}f_{vc}\left[\frac{2}{\omega_{cv}(\omega_{cv}-2\omega)}r^a_{vc}\left(r^b_{vc;c}+r^c_{cv;b}\right)+\frac{1}{\omega_{cv}(\omega_{cv}-\omega)}\left(r^a_{vc;c}r^b_{cv}+r^a_{vc;b}r^c_{cv}\right)+\right.$$
$$\left.\frac{1}{\omega^2_{cv}}\left(\frac{1}{\omega_{cv}-\omega}-\frac{4}{\omega_{cv}-2\omega}\right)r^a_{vc}\left(r^b_{cv}\Delta^c_{cv}+r^c_{cv}\Delta^b_{cv}\right)-\frac{1}{2\omega_{cv}(\omega_{cv}-\omega)}\left(r^b_{vc;a}r^c_{cv}+r^c_{vc;a}r^b_{cv}\right)\right]$$

In which $\left\{r^b_{nv}r^c_{vc}\right\}=(1/2)\left(r^b_{nv}r^c_{vc}+r^c_{nv}r^b_{vc}\right)$ is a symmetrized combination of the dipole matrix elements $r^a_{cn}=\delta_{cn}P^a_{cn}/im\omega_{cn}$, which are in turn obtained from the momentum matrix elements $P^a_{cn}$. Superscripts (a, b, c) refer to the Cartesian coordinates. Here $v$ stands for a valence band state and $c$ for the conduction band state and $n$ the intermediate band, which is either in the valence or in the conduction band. It has been demonstrated by Aspnes [49] that only one virtual-electron transitions (transitions between one valence band state and two conduction band states) give a significant contribution to the second order tensor. Here we have ignored the virtual-hole contribution (transitions between two valence band states and one conduction band state) because it was shown to be negative and more than an order of magnitude smaller than the virtual-electron contribution for these compounds. For simplicity we call $\chi^{(2)}_{abc}(-2\omega;\omega;\omega)$ as $\chi^{(2)}_{abc}(\omega)$. Since the $CuInX_2$ compounds are belong to the point group $\bar{4}2m$ there are only two independent components of the SHG tensor, namely, 123 and 312 components (1, 2, and 3 refer to x, y and z axes, respectively) [50]. These are $\chi^{(2)}_{123}(\omega)$ and $\chi^{(2)}_{312}(\omega)$. In the static limit, these two components are equal according to the Kleninman "permutation" symmetry, which dictates additional relations between tonsorial components beyond the purely crystallographic symmetry.

The second-order nonlinear optical susceptibility is very sensitive to the scissors correction. The scissors correction has a profound effect on magnitude and sign of $\chi^{(2)}_{ijk}(\omega)$. That is attributed to the fact that local density approximation calculations underestimate the energy gaps. Moreover, it is clear from the formulae of $\chi^{(2)}_{ijk}(\omega)$ that, $\chi^{(2)}_{ijk}(\omega)$ depends on the energy gap. Hence when we used the scissors correction we found a considerable effect on $\chi^{(2)}_{ijk}(\omega)$. It is well known that nonlinear optical properties are more sensitive to small changes in the band structure than the linear optical properties. Hence any anisotropy in the linear optical properties is enhanced in the nonlinear spectra. This is attributed to the fact that the second harmonic response $\chi^{(2)}_{ijk}(\omega)$ involves $2\omega$ resonance in addition to the usual $\omega$ resonance. Both $\omega$ and $2\omega$ resonances can be further separated into inter-band and intra-band contributions.

The calculated imaginary part of SHG susceptibility $\chi^{(2)}_{123}(\omega)$ is shown in Fig. 5. We show the $2\omega$ inter-band and intra-band contributions for $CuInX_2$ compounds. We note the opposite signs of the two contributions throughout the frequency range. Both these contributions and the total Im $\chi^{(2)}_{123}(\omega)$ increase on moving from S to Se and decreases on moving from Se to Te. As we note from Figure 1, $CuInSe_2$ and $CuInTe_2$ has a reduction of the energy gap relative to $CuInS_2$. The bandwidth of main groups in band structure and DOS of $CuInSe_2$ and $CuInTe_2$ increases over the value in $CuInS_2$ resulting in a reduction of the gap. Im $\chi^{(2)}_{123}(\omega)$ in$CuInS_2$ is smaller than in $CuInSe_2$ and $CuInTe_2$. From this we conclude that the region around the conduction band minimum does not make a significant contribution to $\chi^{(2)}_{123}(\omega)$. Also the VBM and CBM in $CuInSe_2$ and $CuInTe_2$ seem to be more parallel than in $CuInS_2$. This gives an additional amplifying factor when one takes the integrals over the Brillouin zone, making the oscillator strength of the entire peak larger. This clearly shows that trends in $\chi^{(2)}_{123}(\omega)$ cannot be based only on the minimum band gap. This indicates that the average band gap plays a more significant role.

In Fig. 6 we present the inter-band and intra-band contributions to $\omega$ and $2\omega$ resonances for $CuInTe_2$. We find that $\omega$ resonance is smaller than $2\omega$ resonance. As can be seen the total SHG susceptibility is zero below half the band gap. The $2\omega$ terms start contributing at energies ~1/2 $E_g$ and the $\omega$ terms for energy values above $E_g$. In the low energy regime (≤ eV) the SHG optical spectra is dominated by $2\omega$ contributions. Beyond 3 eV the major contribution comes from $\omega$ terms.

The structures in Im $\chi^{(2)}_{123}(\omega)$ can be understood from the structures in $\varepsilon_2(\omega)$. Unlike the linear optical spectra, the features in the SHG susceptibility are very difficult to identify from the band structure because of the presence of $2\omega$ and $\omega$ terms. But we can make use of the linear optical spectra to identify the different resonance leading to various features in the SHG spectra. The first structure in Im $\chi^{(2)}_{123}(\omega)$ between 1.0 – 3.0 eV for $CuInS_2$, 0.5 – 2.0 eV for $CuInSe_2$, and 1.0 – 2.5 eV for $CuInTe_2$ is associated with interference between a $\omega$ resonance and $2\omega$ resonance and arises from the first hump in $\varepsilon_2(\omega)$. The second structure between 3.0 – 4.5 eV for $CuInS_2$, 2.0 – 3.0 eV for $CuInSe_2$, and 2.5 – 4.5 eV for $CuInTe_2$ is due mainly to $\omega$ resonance and associated

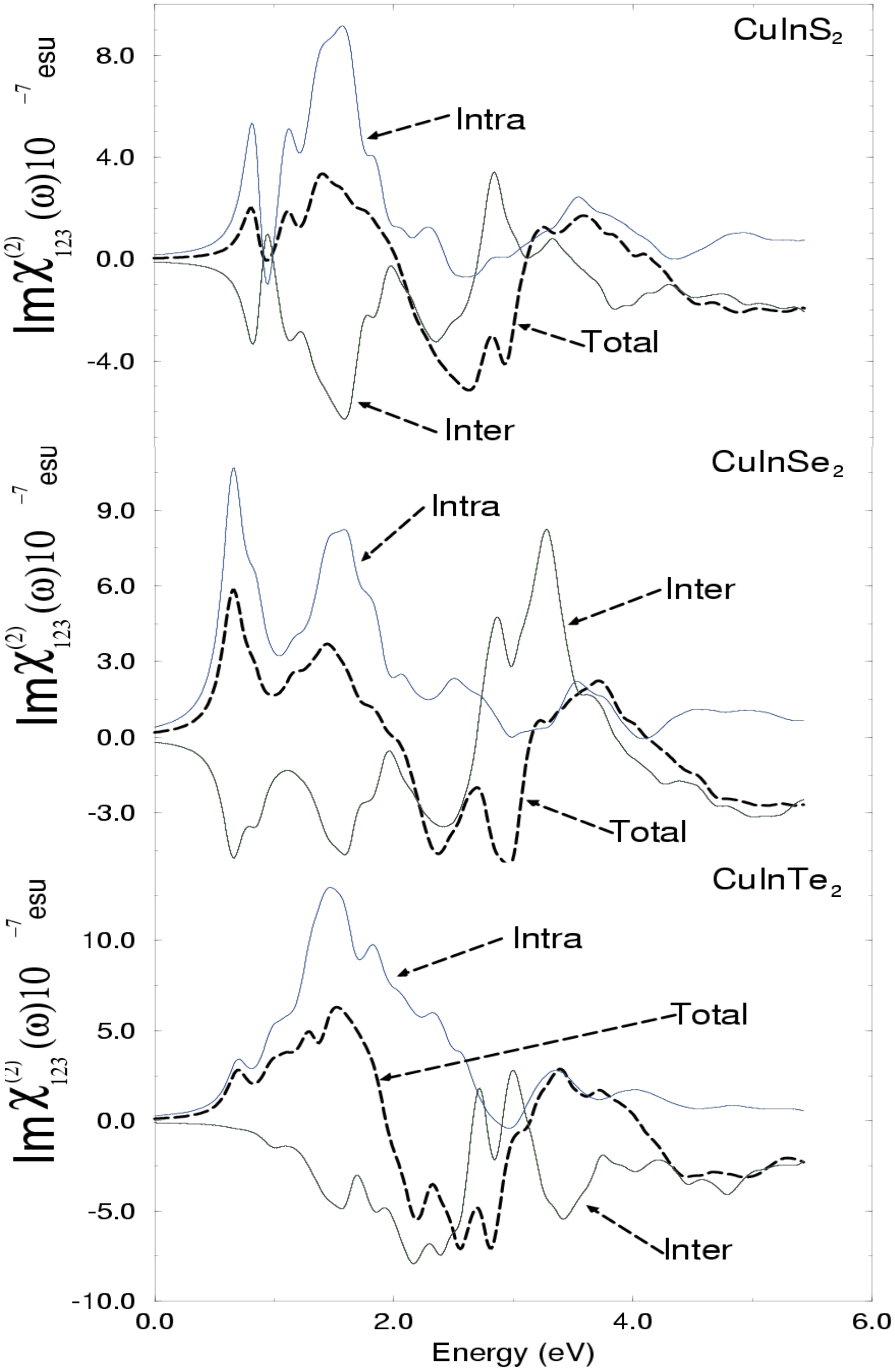

**Figure 5**

Calculated Im $\chi^{(2)}_{123}(\omega)$ along with the ($2\omega$) intra-band and inter-band contributions for $CuInX_2$ compounds.

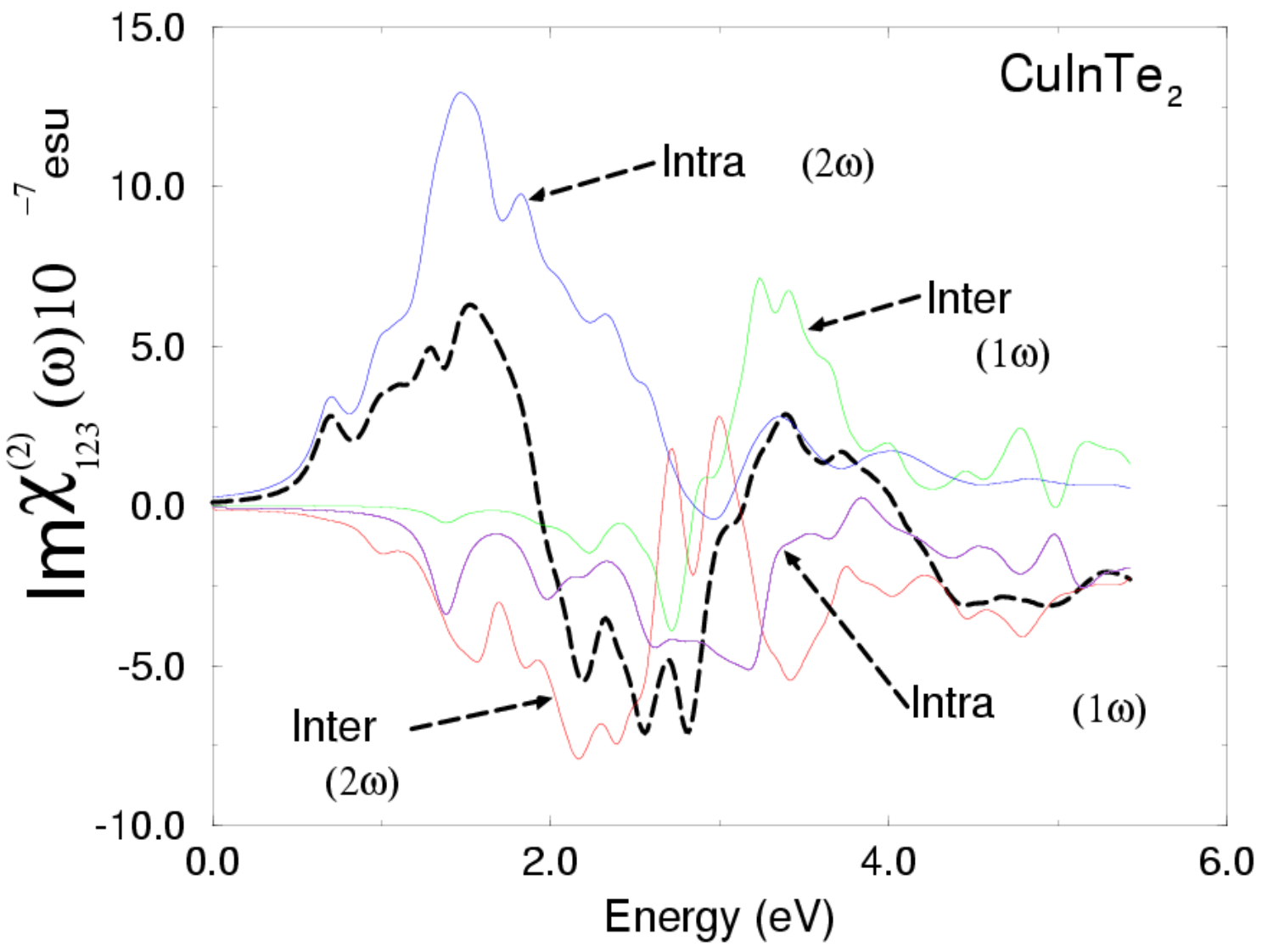

**Figure 6**

Calculated total Im $\chi^{(2)}_{123}(\omega)$, along with the (2$\omega$)/(1$\omega$) intra-band and inter-band contributions for $CuInTe_2$ compound.

---

with high peak in $\varepsilon_2(\omega)$. The last structure from 4.5 – 5.5 for $CuInS_2$, 3.0 – 5.5 eV for $CuInSe_2$, and 4.5 – 5.5 eV for $CuInTe_2$ is manly due to $\omega$ resonance and associated with the tail in $\varepsilon_2(\omega)$.

In Table 2, we present the values of $\chi^{(2)}_{123}(0)$. These values clearly increase on going from S to Se to Te in agreement with experiment and theoretical calculations. We note that there is a large variation in the theoretical and experimental values of $\chi^{(2)}_{123}(0)$, suggesting that these values depend on the method of calculations/measurements. Also some of the calculated vales are equal to the measured one. However our calculated $\chi^{(2)}_{123}(0)$ is in good agreement with some experimental data and some theoretical calculations. For example if we compare our calculated $\chi^{(2)}_{123}(0)$ of $CuInS_2$ with the available experimental data (Table 2) we found that our result appears to be very close to the experimental data [36] whereas the other theoretical value [17] appears to be more than twice the experimental data. This is attributed to our use of the FP. Thus from Table 2, we can conclude that the smaller band gap compounds gives higher values of $\chi^{(2)}_{123}(0)$ in agreement with the experiment [36,51] and theory [17]. The lack of experimental data prevents any conclusive comparison with experiment over a large energy range. It is well known that [60,61] $\chi^{(2)}_{ijk}(\omega) = 2d_{ij}$, based on this expression we compared our calculated $\chi^{(2)}_{312}(0)$ with the values of $d_{36}$ which was obtained by D. Xue et al. [31] at the wavelength 10.6 $\mu$m. Good agreement is found especially both calculations show the negative sign and same trends for both $CuAlSe_2$ and $CuAlSe_2$, and positive sign for the $CuAlTe_2$ (Table 2).

## 4. Conclusion

We have performed calculations of band structure, DOS, frequency dependent linear, birefringence and nonlinear optical response for $CuInX_2$ (X = S, Se, Te) compounds, using FP-LAPW method. Our results for band structure and DOS show that these compounds have similar structures with direct energy band gap. These energy band gap changes when S is replaced by Se/Te. This is attributed to the fact that the bandwidth of conduction bands increases on going from S to Se and decreases on going from Se to Te in agreement with the experimental data. Our calculated energy band gaps show better agreement with the experiment than that obtained by using LMTO method [17]. This is attributed to the use of a FP method. As a result of this the LMTO method is unable to predict correctly the optical transitions, even when the scissor correction is applied. All structures in the imaginary part of the frequency dependent dielectric function $\varepsilon_2(\omega)$ are shifted towards lower energies when S is replaced by Se and Te. We have calculated $\varepsilon_2(\omega)$ and find a considerable anisotropy between $\varepsilon_2^{\perp}(\omega)$ and $\varepsilon_2^{\parallel}(\omega)$. We find that the values of $\varepsilon_1(0)$ increases with decreasing energy gap. This could be explained on the basis of the Penn model. We have calculated the birefringence of these compounds and find it is negative for $CuInS_2$ and $CuInSe_2$ while it is positive for $CuInTe_2$ in agreement with the experimental data. Our calculations of SHG susceptibility show that the intra-band and inter-band contributions are significantly increased when S is replaced by Se/Te. Our calculations show that smaller band gap materials have higher $\chi^{(2)}_{123}(0)$ values. Since $\chi^{(2)}_{123}(0)$ is roughly inversely proportional to the band gap, one might think that this would lead to a possible route to further enhancement of Im $\chi^{(2)}_{123}(\omega)$. The enhancement of SHG is considerable when S is replaced by Se/Te. We have compared our calculated linear and nonlinear optical properties with the previous calculations of D. Xue et al. [31]. Good agreement is found.

**Table 2: Experimental, calculated total and intra-inter- band of the zero frequency of the real part of the $\chi^{(2)}_{123}(\omega)$. $\chi^{123}(0)$ is expressed in units of $1 \times 10^{-8}$ esu.**

| | $CuInS_2$ | $CuInSe_2$ | $CuInTe_2$ |
|---|---|---|---|
| $\chi^{123}(0)$exp. | 14[a], 11[b] | | |
| $\chi^{123}(0)$*total* | 31.7[c], 18* | 72.5[c], 90* | 126.0[c], 135* |
| $\chi^{123}(0)$inter | -51.8[c], -57* | -115.0[c], -110* | -187.2[c], -205* |
| $\chi^{123}(0)$intra | 83.5[c],75* | 187.5[c],200* | 313.2[c], 340* |
| $\chi^{312}(0)$*total* pm/V | -45.3* | -28.2* | 8.5* |
| $d_{36}$ pm/V (Ref.34) | -22.48[d] | -14.51[d] | 3.85* |
| $\chi^{312}(0)$*total* = $2d_{36}$ Pm/V | -44.96** | -29.02** | 7.7** |

[a]Ref.36 [b]Ref.51 [c]Ref.17 [d]Ref.31.
*This work ** The value of $\chi^{123}(0)$*total* (pm/V), calculated from $d_{36}$ (Ref. 31).

## Acknowledgements
The author would like to thank the Institute Computer Center for providing the computational facilities. Also to thank Prof. Claudia Ambrosch-Draxl, 000 Montanuniversität Leoben, 8700 Leoben, Franz-Josef-StraBe 18, Austria, and Dr. Sangeeta Sharma, Institut für Theoretische Physik, Freie Universität Berlin, Arnimallee 14, D-14195 Berlin, Germany. This work was supported from the institutional research concept of the Institute of Physical Biology, UFB (No. MSM6007665808), and the Institute of System Biology and Ecology, ASCR (No. AVOZ60870520).